\documentclass[pre,aps,twocolumn,showpacs,floatfix]{revtex4}
\usepackage[dvips]{graphicx}
\usepackage{amsmath}
\usepackage{amssymb}
\usepackage[nice]{nicefrac}

\begin{document}
\title{Anomalous diffusion and generalized Sparre-Andersen scaling}

\author{Bart{\l}omiej Dybiec}
\email{bartek@th.if.uj.edu.pl}
\affiliation{Marian Smoluchowski Institute of Physics, and Mark Kac Center for Complex Systems Research, Jagellonian University, ul. Reymonta 4, 30--059 Krak\'ow, Poland}

\author{Ewa Gudowska-Nowak}
\email{gudowska@th.if.uj.edu.pl}
\affiliation{Marian Smoluchowski Institute of Physics, and Mark Kac Center for Complex Systems Research, Jagellonian University, ul. Reymonta 4, 30--059 Krak\'ow, Poland}

\date{\today}
\begin{abstract}

We are discussing long-time, scaling limit for the anomalous diffusion composed
of the subordinated L\'evy-Wiener process.
The limiting anomalous diffusion is in general non-Markov, even
in the regime, where ensemble averages of a mean-square displacement or
quantiles representing the group spread of the distribution follow the scaling
characteristic for an ordinary stochastic diffusion. To discriminate between
truly memory-less process and the non-Markov one, we are analyzing deviation of
the survival probability from the (standard) Sparre-Andersen scaling.
\end{abstract}

\maketitle

%
%
\section{Introduction}
The question of the time that it takes for stochastic process to reach a
specific point or state by the first time is central in many applications of
stochastic modeling in physics (Kramers problem \cite{vankampen1981}), chemistry
(reaction kinetics \cite{redner2001}), biology (neural activity models
\cite{goelrichter1974}) and economics (estimation of the ruin time
\cite{feller1968}). For a random walk sequence, a nontrivial theorem due to
Sparre-Andersen \cite{sparre1953,sparre1954} states that the asymptotic form of
the probability $S_n(x_0)$ of not crossing the boundary within the first $n$
steps after starting the motion at $x_0 > 0$ (otherwise named the survival
probability on the positive semi-axis) does not depend on the form of a jump length
distribution if only it is symmetric, continuous and Markovian. For a large number of steps,
one invariably has regardless of the exact jump length distribution type) $S_n(x_0) \propto c(x_0)n^{-1/2}$, where the prefactor
$c(x_0)$ depends on the initial position. The result can be easily generalized
for the continuous-time version of the process. For unbiased, continuous
Gaussian random walk its first passage time density (FPTD) from $x_0$ to $x$ can
be easily calculated explicitly \cite{redner2001}: Let us denote
$p(x_2,t|x_0,0)=p(x_2-x_0)$ the
probability of motion from a position $x_0$ to $x_2$ in time $t$ with $x$
denoting the position on the way from $x_0$ to $x_2$, i.e. $x_0<x <x_2$. By
taking $f(x,\tau|x_0,0)=f(x-x_0,\tau)$ as the probability of arriving for the first time at $x$
at time $\tau$, the equation for $p(x_2-x_0,t)$ reads
\begin{equation}
p(x_2-x_0,t) = \int^t_0p(x_2-x, t-\tau)f(x-x_0,\tau) d\tau
\end{equation}
with its Laplace transform given by
\begin{equation}
\tilde{p}(x_2-x_0,s) = \tilde{p}(x_2-x, s)\tilde{f}(x-x_0,s).
\end{equation}
For unbiased Gaussian random walk we have $p(x,t|0,0)=p(x,t)=(2\pi\sigma^2
t)^{-1/2}\exp(-\frac{x^2}{2\sigma^2t})$. With the Laplace transform
$\tilde{p}(x,s)=\int^{\infty}_0 p(x,t) e^{-st}
dt=(2s\sigma^2)^{-1/2}\exp(-\sqrt{2x^2s/\sigma^2})$ one obtains
\begin{equation}
\tilde{f}(x-x_0,s)=\exp(-\sqrt{2(x-x_0)^2s/\sigma^2})
\end{equation}
which, by inverse Laplace transform, yields the L\'evy-Smirnov distribution
\begin{equation}
f(x,t|x_0,0)=f(x-x_0,t)=\frac{1}{t}\sqrt{\frac{(x-x_0)^2}{2\pi \sigma^2t}}\mathrm{e}^{-\frac{(x-x_0)^2}{2\sigma^2 t}},
\label{eq:l-s}
\end{equation}
where $x_0$ represents the initial condition.
This ``inverse Gaussian distribution'' decays for long times as $f(x,t|x_0,0)
\propto t^{-3/2}$ and does not have a first moment, i.e. the mean first passage time from $x_0$ to $x$
diverges. On the other hand, since $\int_0^{\infty}f(x,t|x_0,0)dt=1$, the
particle performing the one dimensional Gaussian random walk will certainly hit any point $x$ during its motion.

Assuming the absorbing boundary located at the origin, i.e. at $x=0$, formula
(\ref{eq:l-s}) with $x=0$ gives the probability density of the first passage
time from the positive semi-axis for a Gaussian random walk.
It should be stressed, however that for generally non-Gaussian noises, the
knowledge of the boundary location may be insufficient to specify in full the
corresponding conditions for absorption or reflection
\cite{dybiec2006,metzler2004,zoia2007,koren2007b}. In particular trajectories of
L\'evy walks may exhibit discontinuous jumps and in a consequence, the location
of the boundary itself is not hit by the majority of sample trajectories. In
order to properly take care of possible excursion of the trajectories beyond the
location of the boundary (at, say $x=0$) with subsequent re-crossings into the
interval ($x>0$), the whole semi-line ($x\leqslant 0$) has to be assumed
``absorbing''. This nonlocal definition of the boundary conditions secures
proper evaluation of the first passage time distribution and survival
probability \cite{dybiec2006,metzler2004,zoia2007,koren2007b}, see below.

The very same scenario, see Eq.~(\ref{eq:l-s}), dictated by the
Sparre-Andersen theorem holds also true
for ``paradoxical'' diffusion-like processes studied in terms of CTRW
(continuous time random walks) where kinetics of the walker is determined by the
distribution of jump lengths and distribution of waiting times before a next
jump to occur \cite{sokolov2004b}. If the process is regular in time but with
nontrivial jump distribution following the (symmetric) L\'evy law of stability
(so called symmetric L\'evy flight), the first passage time density (FPTD)
follows the Sparre-Andersen universality \cite{metzler2004,chechkin2006,dybiec2006}.
Notably, however, if subordinating the number of steps $n$ to the physical
clock time $t$ such that the number of steps $n$ per unit of physical time
follows some distribution with a power-law tail $p(n, t=1) \propto n^{-(1+\beta)}$
with $0<\beta\leqslant 1$, the deviations from the universality can be observed
\cite{sokolov2004b}.

To further elucidate the nature of deviation from the ``standard'' Sparre-Andersen scaling in
subordinated scenarios, we consider the process $X(t)=\tilde{X}(S_t)$, for
which the parental process $\tilde{X}(\tau)$ is described by a Langevin equation
\cite{magdziarz2007b}
\begin{equation}
dX(\tau)=\sigma dL_{\alpha}(\tau)
\label{eq:def}
\end{equation}
driven by a symmetric $\alpha$-stable L\'evy motion $L_{\alpha}(\tau)$ with the
Fourier transform $\langle e^{ikL_{\alpha}(\tau)}
\rangle=e^{-\tau|k|^{\alpha}}$. Here $\tau$ stands for the operational time
scale which is changed to the physical time scale $t$ by subordination
$\tilde{X}(S_t)$. The subordinator $S_t$ is defined as $S_t=\mathrm{inf}\left
\{\tau: U(\tau)>t\right \}$ with $U(\tau)$ denoting a strictly increasing
$\nu$-stable L\'evy motion ($0<\nu<1$) and is assumed to be independent from the noise
term $L_{\alpha}(\tau)$.

The above setup has been recently proved
\cite{magdziarz2007b,magdziarz2008,magdziarz2007} to give a proper stochastic
realization of the random process described otherwise by a fractional diffusion
equation
\cite{saichev1997,jespersen1999,metzler1999,metzler2000,sokolov2002,
metzler2004}
\begin{equation}
 \frac{\partial p(x,t)}{\partial t}={}_{0}D^{1-\nu}_{t}\left[ \sigma^\alpha
\frac{\partial^\alpha}{\partial |x|^\alpha} \right] p(x,t).
\label{eq:ffpe}
\end{equation}
Here ${}_{0}D^{1-\nu}_{t}$ denotes the Riemann-Liouville fractional derivative
defined as
${}_{0}D^{1-\nu}_{t}f(t)=\Gamma(\nu)^{-1}\frac{d}{dt}\int^t_0(t-\tau)^{\nu-1}
f(\tau)d\tau$ with $0<\nu\leqslant 1$
 and
$\frac{\partial^\alpha}{\partial |x|^\alpha} $ with $0<\alpha\leqslant 2$ stands
for the Riesz fractional derivative with the Fourier transform
$\mathcal{F}[\frac{\partial^\alpha}{\partial |x|^\alpha}
f(x)]=-|k|^{\alpha}\hat{f}(k)$ \cite{podlubny1998}. Occurrence of the operator ${}_{0}D^{1-\nu}_{t}$ is due to the heavy-tailed waiting times between successive jumps and presence of the Riesz fractional derivative $\frac{\partial^\alpha}{\partial |x|^\alpha} $ is a consequence of the L\'evy-flight character of the jumps.

In this paper, instead of seeking an analytical solution to Eq.~(\ref{eq:ffpe}),
we switch to a Monte Carlo method
\cite{magdziarz2007b,magdziarz2008,magdziarz2007,gorenflo2002,meerschaert2004}
which allows generating trajectories of the subordinated process $X(t)$ with the
parent process $\tilde{X}(\tau)$. Furthermore, we study the potential free case,
see Eq.~(\ref{eq:ffpe}), i.e. we assume $V(x)$=0. The assumed algorithm provides
means to investigate the competition between subdiffusion (controlled by
$\nu$-parameter) and L\'evy flights characterized by a stability index $\alpha$.

For Markov processes, the Sparre-Andersen scaling \cite{sparre1953,sparre1954} presents a universal law which is  independent of detailed properties of the jump length distribution (if it is only continuous and symmetric).  In particular, for continuous times, the scaling predicts the  $t^{-1/2}$ decay of the survival probability, independently of whether the moments of the underlying jump process exist or not. For example, for $\alpha<2$, the moments of the process $X(t)$ (cf. Eq.~(\ref{eq:def})) exist only for $\delta<\alpha$ with obvious divergence of moments of order
$\delta\geqslant\alpha$, i.e.
\begin{equation}
\langle |X|^\delta \rangle = \int\limits_{-\infty}^\infty |x|^\delta p(x,t) dx = \infty.
\end{equation}
This divergence can be easily demonstrated in the case of (pure) L\'evy flights described by Eq.~(\ref{eq:def}), for which the operational time $\tau$ and physical time $t$ is equivalent, i.e. $S_t=t$ and consequently $t=\tau$, see below.
In such a case, the $p(x,t)$ is a L\'evy stable density (whose width is growing with time) and $\langle |X|^\delta \rangle$ stays infinite for $\delta\geqslant \alpha$.
Clearly, for finite time $t$ and finite number of representative trajectories $N$ (otherwise called realizations of the process $X(t)$), variance $\langle X^2 \rangle$ of (symmetric) L\'evy flights
stays finite, see \cite[Eq.~(1.19)]{bouchaud1990} and \cite{dybiec-anomalous}.
In fact, finite number of realizations (to be distinguished from the number of steps $n$ used in simulation of a single trajectory of time duration $t=n\Delta t$) and finite time introduce an
effective cutoff  to the jump length distribution.
In contrast,  for any fixed time the variance diverges with increasing number of simulated trajectories $N$.
Analogously, for any fixed $N$, the variance diverges with increasing time (scaling like $t^{1/\alpha}$, see \cite[Eq.~(1.19)]{bouchaud1990} and \cite{dybiec-anomalous}).
The problem of mathematical divergence can be resolved either by introducing  spatiotemporal coupling (typical for so called L\'evy walks \cite{shlesinger1986,blumen1989}) or by proper truncation of the jump length
distribution \cite{mantegna1994b,koponen1995,delcastillonegrete2008b,sokolov2004}. Several truncation methods have been proposed \cite{mantegna1994b,koponen1995,delcastillonegrete2008b,sokolov2004} to  retain the finite second moment.
In particular, paralleling the simulation studies of Mantegna and Stanley \cite{mantegna1994b}, a smooth exponential cutoff has been introduced by Koponen \cite{koponen1995}.
Instead of truncating
tails of a distribution, this approach is based upon the exponential tempering
of the L\'evy density and preserves  the infinite-divisibility \cite{feller1968}
of the distribution. The classical tempered stable distribution has been further generalized by Rosi\'nski (for more detailed discussion see \cite[Chapter 5.7]{klages2008} and \cite{koponen1995,chechkin2008,rosinski2007}).

%
%
\section{Methods \& Results}

For systems driven by symmetric process the generalized Sparre-Andersen scaling
\cite{sparre1953,sparre1954,feller1968,redner2001,koren2007c} can be used to discriminate
between Markovian and non-Markovian situations. More precisely, according to the
Sparre-Andersen theorem for a stochastic processes driven by any symmetric white
noises, the first passage time densities, $f(t)=\frac{d{F}}{dt}$, from the real
half line asymptotically behave like
\begin{equation}
f(t) \propto t^{-3/2}.
\end{equation}
Consequently the survival probability $S(t)$, i.e. the probability of finding a
particle starting its motion at $x_0>0$ in the real (positive) half line, scales
like
\begin{equation}
 S(t)=1-{F}(t) \propto t^{-1/2}.
\label{eq:sparre}
\end{equation}
Therefore, any deviation of the survival probability from $t^{-1/2}$ dependence
indicates violation of assumptions assuring the proof of the theorem. It can
mean either that a system is driven by non-symmetric or not ``memory-less''
driving. In consequence, for symmetric drivings, analysis of data based on
(assumed a priori) Sparre-Andersen scaling may reveal deviations from
Markovianity.

We study statistical properties of a symmetric free L\'evy motion
Eq.~(\ref{eq:def}) constrained to the initial position $x(0)=1$. To achieve the
goals, we use the scheme of stochastic subordination
\cite{magdziarz2007b,magdziarz2008,magdziarz2007,dybiec-anomalous}, i.e.
we obtain the process of primary interest $X(t)$ as a function
$X(t)=\tilde{X}(S_t)$ by randomizing the time clock of the process $X(\tau)$ using
a different clock $S_t$. The parent process $\tilde{X}(\tau)$ is composed of
increments of symmetric $\alpha$-stable motion described in an operational time
$\tau$ and in every jump moment the relation $U(S_t)=t$ is fulfilled. The
(inverse-time) subordinator $S_t$ is (in general) non-Markovian hence, as it
will be shown, the diffusion process $\tilde{X}(S_t)$ possesses also some degree
of memory.

The survival probability, see Eq.~(\ref{eq:sparre}), was estimated from
ensemble of trajectories of the process $X(t)$ starting at $x_0$ ($x_0>0$).
For $\alpha<2$, in order to correctly account for non-local boundary conditions \cite{metzler2004,dybiec2006,zoia2007} we have excluded multiple recrossing events, i.e. every time the particle reached any point
$x$ beyond the boundary it was removed from the system.

\begin{figure}[!ht]
\begin{center}
\includegraphics[angle=0,width=0.9\columnwidth]{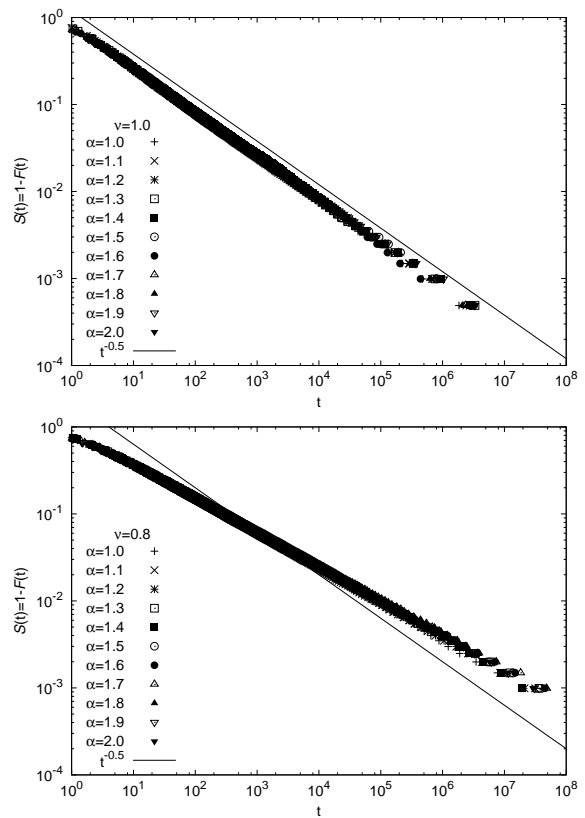}
\caption{Sparre Andersen scaling (\ref{eq:sparre}) for $\nu=1$ (top panel) and
$\nu=0.8$ (bottom panel) with various $\alpha$. The process was numerically
approximated by subordination techniques with $\Delta t=10^{-2}$ and averaged
over $N=10^5$ realizations, $\sigma=1/\sqrt{2}$.}
\label{fig:sa_fixed_sub}
\end{center}
\end{figure}

\begin{figure}[!ht]
\begin{center}
\includegraphics[angle=0,width=0.9\columnwidth]{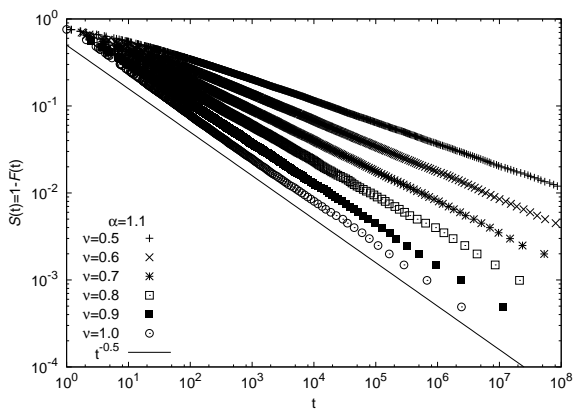}
\caption{Sparre Andersen scaling (\ref{eq:sparre}) for $\alpha=1.1$ with various
$\nu$. Simulation parameters as in Fig.~\ref{fig:sa_fixed_sub}. In
Figs.~\ref{fig:sa_fixed_sub} -- \ref{fig:sa_fixed_alpha} initial position was
set to $x(0)=1$. However, due to the Sparre-Andersen theorem, results with
other values of $x(0)$ are perfectly coherent with results for $x(0)=1$ (not
shown) and lead to the same values of exponent $b$, see
Eq.~(\ref{eq:exponent}).}
\label{fig:sa_fixed_alpha}
\end{center}
\end{figure}

In Figs.~\ref{fig:sa_fixed_sub}--\ref{fig:sa_fixed_alpha} the survival
probability $S(t)=1-{F}(t)$ is depicted for various stability indices $\alpha$
and various subdiffusion parameters $\nu$. It is clearly visible that the
survival probability $S(t)$ behaves like a power-law for all considered values
of the subdiffusion parameter $\nu$ and stability index $\alpha$. However, the
exponent characterizing the power-law dependence is equal to $-1/2$, as
predicted by the (standard) Sparre-Andersen theorem, only for the Markovian case ($\nu=1$).
In more general case the power-law is characterized by the exponent $b$
\begin{equation}
S(t)=1-{F}(t) \propto t^{b}
\label{eq:exponent}
\end{equation}
which differs from
\begin{equation}
b=-\frac{1}{2}.
\end{equation}
For $\alpha=2$ with any $\nu$ ($0<\nu\leqslant 1$), the first passage time distribution is one sided L\'evy
distribution characterized by the stability index $\nu/2$
\cite{Balakrishnan1985, metzler2000b,barkai2001}, i.e.
\begin{equation}
 b=-\frac{\nu}{2}.
\end{equation}
Furthermore, in the general case, the value of the exponent $b$ does not depend
on the stability index $\alpha$ of the jump length distribution
\cite{koren2007c}.

\begin{figure}[!ht]
\begin{center}
\includegraphics[angle=0,width=0.9\columnwidth]{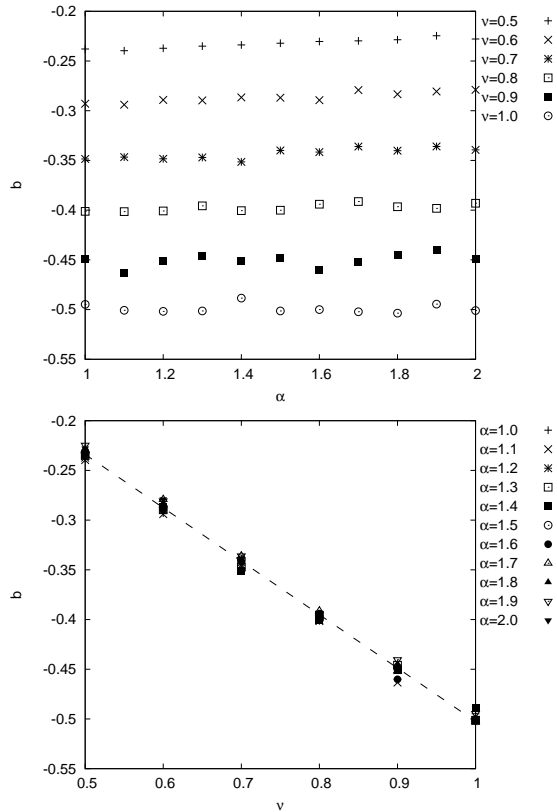}
\caption{Exponent $b$, see Eq.~(\ref{eq:exponent}), characterizing power-law
behavior of the survival probability $S(t)$ as a function of the stability index
$\alpha$ (top panel) and subdiffusion parameter $\nu$ (bottom panel). Simulation
parameters as in Fig.~\ref{fig:sa_fixed_sub}.}
\label{fig:exponents}
\end{center}
\end{figure}

Figs.~\ref{fig:sa_fixed_sub}--\ref{fig:sa_fixed_alpha} confirm that the
value of
the exponent $b$ depends on the subdiffusion parameter, $\nu$, only.
Fig.~\ref{fig:sa_fixed_alpha} shows results for $\alpha=1.1$. Results for others
values of stability index $\alpha$ are the same as those one for $\alpha=1.1$.
Finally, Fig.~\ref{fig:exponents} presents value of the exponent $b$, see
Eq.~(\ref{eq:exponent}), as the function of the subdiffusion parameter $\nu$ and
stability index $\alpha$. Fig.~\ref{fig:exponents} confirms that exponent $b$
depends on the subdiffusion parameter $\nu$ and the influence of the stability
index $\alpha$ is negligible. Furthermore, $b$ depends linearly on
$\nu$: $b=-(0.54\pm 0.01)\nu+(0.03 \pm 0.03)$, what agrees with earlier findings
\cite{Balakrishnan1985,barkai2001,koren2007c}, see
Fig.~\ref{fig:exponents}. Value of the exponent $b$ is the decreasing function
of the subdiffusion parameter $\nu$ leading to the slowest decay of the survival
probability for small values of the $\nu$ parameter, i.e. when the exponent
$\nu$ deviates the most from its Markovian -- ``memory-less'' value -- 1. The
deviation of the exponent $b$ from $-1/2$ clearly indicates a typical slowing
down of the subdiffusive process in comparison to its (Markov) regular diffusion
analogue. The $\alpha$-independence of the survival probability $S(t)$ in this
case shows that the properties of the decay kinetics are determined by the
subdiffusive part of the process only. This observation is different from the
results obtained by Sokolov and Metzler for a class of L\'evy random processes
subordinated (via the relation connecting distribution of number of jumps $n$ in
physical time $t$) to L\'evy flights or to Brownian random walks. In particular,
in their derivation of subordination, the Authors are using the Markovian L\'evy
flight process $X(t)$ transformed to the process $X(T(t))$ by use of the
operational time $T$ which, by itself, is called the directing process $T(t)$.
The density for the process $X(T(t))$ assumes the form $P(x,t)=\int^{\infty}_0
p(x, \tau)r(\tau,t)d\tau$ with $p(x,t)$, $r(\tau, t)$ representing densities of
a L\'evy flight process and the density of the directing process, respectively.
If $X(t)$ is a stable process with a stability parameter $\alpha$ and $T(t)$ is
a one-sided stable process with exponent $\nu$, the subordinated process
$X(T(t))$ becomes a stable process with the stability index $\nu\alpha$. In
contrast, in more general terms of the CTRW scenario, after waiving the
assumption about independent increments of the $T(t)$ process, the asymptotic
form of the distribution $P(x,t)$ can be derived by use of Tauberian theorems
\cite{saichev1997} and is known to be $\nu/\alpha$ self-similar, i.e.
$P(x,t)=t^{-\nu/\alpha}P(xt^{-\nu/\alpha},1)$ \cite{magdziarz2007,dybiec2009}.

%
%
\section{Summary \& Conclusions}

We have discussed effects of the subordination scheme leading to the fractional
diffusion equation Eq.~(\ref{eq:ffpe}).
By use of the Monte Carlo method we have created trajectories of the process
$X(t)=\tilde{X}(S_t)$ with $S_t$ being the inverse time $\alpha$-stable
subordinator. Since the $S_t$ process appears as an asymptotic one in the CTRW
scheme with heavy-tailed waiting time distribution between successive jumps and
the parental process $X(\tau)$
is assumed symmetric $\alpha$-stable, the proposed subordination
\cite{magdziarz2007} leads to $
\nu/\alpha$ self-similar process whose survival probabilities are governed by the
stability exponent $\nu$. Information gained from the analysis of generated
trajectories brings around further confirmation of non-Markov property of the
motion \cite{dybiec-anomalous}. Moreover, due to the interplay between the
subdiffusion in time and superdiffusion in step lengths, the resulting process
violates the ergodicity (in the weak sense) so that the long time average is
different from the average taken over the ensemble of trajectories
\cite{rebenshtok2007,rebenshtok2008,dybiec2009b}. This issue is of special interest in the
context of single-particle measurements \cite{metzler2009} which require
analysis of time series representative for the motion. In this work we
demonstrate that subdiffusive and non-Markovian character of the motion can be
grasped by analyzing survival probabilities which deviate from the
(standard) Sparre-Andersen scaling also in those cases when the ensemble averages suggest a
Brownian diffusion with $\nu/\alpha=1/2$ \cite{dybiec-anomalous}.

%
%
\acknowledgments
The research has been supported by the Marie Curie TOK COCOS grant (6th EU
Framework Program under
Contract No. MTKD-CT-2004-517186). Additionally, BD
acknowledges the support from the Foundation for Polish Science.

%

\end{document}